\documentclass[aps,prd,nofootinbib,twocolumn,groupedaddress]{revtex4}
\usepackage{graphicx}

\begin{document}


\title{Toward a Unified Description of Dark Energy and Dark Matter from the Abnormally Weighting Energy Hypothesis}


\author{A. F\"uzfa$^{1}$\footnote{F.N.R.S. Postdoctoral Researcher, Associated Researcher at LUTh, Observatory of Paris and at Center for Particle Physics and Phenomenology (CP3), Universit\'e Catholique de Louvain (UCL)}}
\author{J.-M. Alimi$^{2}$\footnote{Associated Researcher at GAMASCO, FUNDP}}
\affiliation{$^{1}$Groupe d'Application des MAth\'ematiques aux Sciences du COsmos (GAMASCO), 
University of Namur (FUNDP), Belgium\\
$^2$Laboratory Universe and Theories (LUTh), CNRS UMR 8102,\\
 Observatoire de Paris-Meudon and Universit\'e Paris VII, France}


\date{\today}

\begin{abstract}
The Abnormally Weighting Energy (AWE) hypothesis consists of assuming that the dark sector of cosmology
violates the weak equivalence principle (WEP) on cosmological scales, which implies a violation of the strong equivalence principle for ordinary matter. 
In this paper,
dark energy (DE) is shown to result from the violation of WEP by pressureless (dark) matter.
This allows us to build a new cosmological framework
in which general relativity (GR) is satisfied at low scales, as WEP violation depends on the ratio of the ordinary matter over dark matter densities, but 
at large scales, we obtain a general relativity-like theory with a different
value of the gravitational coupling.
This explanation is formulated in terms of a tensor-scalar theory of
gravitation without WEP for which there exists a revisited convergence mechanism toward GR. 
The consequent DE mechanism build upon the anomalous gravity of dark matter 
(i) does not require any violation of the strong energy condition $p<-\rho c^2/3$, 
(ii) offers a natural way-out of the coincidence problem thanks to the non-minimal couplings to gravitation, (iii)
accounts fairly for supernovae data from various very simple couplings and with density parameters very close to the
ones of the concordance model $\Lambda CDM$, therefore suggesting an explanation to its remarkable adequacy. Finally, (iv) this mechanism ends up in the future with an Einstein-de Sitter expansion regime once
the attractor is reached. 
\end{abstract}


\maketitle

\section{Introduction}
In the past decade, cosmology has entered into an era of high precision. 
Large observational projects like sky surveys (SDSS \cite{sdss}, 2dF \cite{2DF}, ...), precise measurements of the cosmic microwave background (CMB) 
anisotropies (COBE \cite{cobe}, BOOMERanG \cite{boomerang} and WMAP \cite{wmap}) and the extensive use of type Ia supernovae to characterize the recent cosmic 
expansion (SNLS \cite{snls}, The Supernovae Cosmology Project \cite{scp}, High-z SN search \cite{highz}) have not only achieved to settle 
the hot Big Bang scenario as the basic framework to describe the whole universe but also have completely changed our past vision of its energy
content. They have indeed revealed the recent domination of an unexpected and still unexplained energy contribution, the so-called dark energy (DE), 
which accelerated the recent cosmic expansion. Therefore, if the existence of DE can 
hardly be contested nowadays due to the accumulating observational evidence, 
the question of its physical origin has become a crucial problem not only for theoretical cosmology but also for fundamental physics.
\\
\\
In this paper, we apply the Abnormally Weighting Energy (AWE) Hypothesis, introduced in \cite{fuzfa}, to 
a pressureless fluid (our "\textit{dark matter}" -- DM -- here). This allows not only accounting for
the observed cosmic acceleration but also offers fascinating perspectives about several DE problems like
alternative to negative pressures, cosmic coincidence, relations between DM and DE and the fate of the Universe.
The AWE hypothesis consists of assuming that DE
does not couple to gravitation in the same way as usual matter (baryons, photons, etc.).
This implies a violation of the weak equivalence principle (WEP),
mostly on cosmological scales, where DE dominates the energy content of the Universe. 
This anomalous weight also implies that the related gravitational binding
energy of DE does not generate the same amount of gravity than usual matter, yielding
a violation of the strong equivalence principle (SEP). This will result in a running gravitational coupling
constant on cosmological scales whose dynamics reproduce the observed cosmic acceleration.\\
\\
In previous works \cite{fuzfa}, we have applied the AWE hypothesis to a Born-Infeld gauge interaction to show how
the Hubble diagram of type Ia supernovae can be explained in terms of both cosmic acceleration and variation
of the Newton gravitational constant on small scales. Here, we show that the above results actually do not depend on
a particular equation of state of the AWE sector. Furthermore, we also provide
a detailed analysis of the AWE dynamics that shows
how cosmic acceleration is a generic prediction of the AWE hypothesis resulting from the competition between different non-minimal couplings.
We also establish that the attractor to which gravitation is driven depends on the ratio between the energy densities of 
ordinary matter and AWE Dark Matter. Therefore, General Relativity (GR) can be retrieved on small scales 
where the structure formation has largely favoured the abundance of normally weighting matter like baryons. 
\\
\\
Of course, the idea that DM violates the equivalence principle has been suggested for a long time (see \cite{pscora} and references therein).
Usually, this violation results from the variation
of the inertial mass of DM particles which is ruled by massive (yet ultra-light) quintessence scalar field (DM-DE interactions). These models therefore rely
on a self-interaction potential of the scalar field and a violation of the strong energy condition (SEC, see \cite{caroll})
$p<-\rho/3$  to provide the necessary cosmic acceleration. 
Nevertheless, the DM-DE couplings
enhances the last: it is shown in \cite{pscora} that 
the interaction of DM with quintessence can be interpreted as a ghost equation of state $\omega_{eff}=p/\rho<-1$ in the recent past of cosmic history.
In the present models, we will show how such a phantom equation of state can be easily obtained in the observable frame.
Moreover, the usual DM-DE coupling considered in \cite{pscora} and references therein assumes a negligible coupling between quintessence and baryons
which is, somehow, an even worst violation of the WEP than what is assumed here: why should the coupling strengths to ordinary matter vanish while
those of the hidden sector are non-negligible? In the AWE hypothesis, they are both assumed to be of order unity, which is crucial for obtaining
a cosmic acceleration without a self-interaction potential of the scalar field.
In \textit{chameleon cosmology} \cite{chameleon} however, 
both baryons and DM are non-minimally coupled to the quintessence field, rendering the mass of the last density-dependent. 
This offers the possibility of accounting for the present bounds of local tests of SEP and WEP on Earth and in the solar system by the higher mass of the scalar field
in a denser neighbourhood than in the cosmological limit.
As in the present work, the amplitude of the non-minimal couplings to the scalar field in \cite{chameleon} are also of order unity.
The AWE hypothesis can be seen also as a generalization of chameleon fields in three ways. (i) It makes the most of the different non-minimal couplings. Indeed, the stabilization of this massless scalar field in the minimum of an effective potential is obtained through the competition
of the different non-minimal couplings of AWE and ordinary matter and not through the competition between self-interaction potential and non-minimal coupling to DM like in \cite{chameleon}.
(ii) We take
into account more general non-minimal couplings to a massless scalar field. Finally (iii), the observational quantities will be here written in terms of the metric coupling universally to usual
matter, which allows reducing cosmic acceleration to a frame effect. 
\\
\\
This paper is structured as follows. In the next section, the AWE hypothesis is introduced as a natural generalization of several DE models
before being applied to a pressureless fluid. 
In section III, we describe observable properties of the AWE cosmology while in section IV, we show how the anomalous gravity of DM can account successfully for the 
present state of expansion of the Universe.
Several key questions related to
DE like coincidence problem, influence of parametrization, 
values of density parameters and the fate of the Universe are discussed before we conclude in section V.  
In the appendix, the reader will find the details concerning the 
convergence mechanism toward GR revisited by the AWE Hypothesis.
\section{The Abnormally Weighting Energy (AWE) Hypothesis}
\subsection{The Dark Cosmology Iceberg}
The usual explanation to account for the observational evidence presented before is to reintroduce the long-time discarded cosmological constant $\Lambda$, 
which has been historically introduced by Einstein himself in 1917 as a Mach principle-inspired term \cite{einstein}. 
The corresponding gravitational theory derives from the well-known Einstein-Hilbert
action of GR:
\begin{equation}
\label{eh}
S_{EH}=\frac{1}{2\kappa_*}\int\sqrt{-g_*}d^4x\left\{R^*+2\Lambda\right\}+S_m\left[\psi_m,g^*_{\mu\nu}\right],
\end{equation}
where $g^*_{\mu\nu}$ is the Einstein metric, $\Lambda$ the cosmological constant, $\kappa_*=8\pi G_*$ is the bare gravitational coupling constant, $R^*$ is the curvature scalar and $\psi_m$ are the matter fields.
Reintroducing the cosmological constant therefore does not require to either modify general relativity, as it is part of it, nor the cosmological principle stating 
the homogeneity and isotropy of the universe on large-scales (for interpretation of DE based on
inhomogeneity effects see \cite{buchert} and references therein). The resulting cosmological model, called the concordance model $\Lambda CDM$, 
has only one additional parameter to fully determine the DE sector: the value of the cosmological constant $\Lambda$. \\
\\
Although this situation might look satisfactory from an observational point of view,
the situation is however worst from the theoretical side (see \cite{weinberg} for an introduction to the cosmological
constant problems). First, the effect of a (positive) cosmological constant on 
gravitation is similar to a fluid with negative pressure, producing the necessary cosmic acceleration. 
This reminds quantum fluctuations around the vacuum state in quantum field theory. 
Unfortunately, the computation of the value of the vacuum energy by integrating all zero-point energies
up to the Planck scale where general relativity should enter its quantum regime 
leads to the huge value of $\rho_\Lambda^{th}\approx m_{Pl}^4\approx 10^{76} GeV^4$ (in geometrical units where $\hbar=c=1$, $G=m_{Pl}^{-2}$). 
What is worst now with the astrophysical
evidence for the cosmological constant, is that the vacuum energy density is indeed not vanishing but is of cosmological order of magnitude:
$\rho_{\Lambda}\approx \rho_{c}(z=0)\approx 10^{-47} GeV^4$.
The consequent overestimation 
of the cosmological constant is about 120 orders of magnitude with regards to its observed value. 
This constitutes the so-called fine-tuning problem: which divinely precise mechanism can reduce 
this estimation of the cosmological constant by such a gigantic amount?
From a theoretical physics point of view, there is yet no satisfactory reason as to why the vacuum energy vanishes, despite several attempts 
(see \cite{weinberg}).
Therefore, in any alternative modeling of dark energy,
this problem will be ignored and some yet undiscovered mechanism or symmetry is invoked
to cancel out the huge value of the cosmological constant. 
In addition to this fine-tuning problem, there is also another puzzling question: 
why is the vacuum energy density, something that should be set once for all in the universe and should depend on quantum mechanics, 
precisely of the order of the critical density today? 
Why is this quantity of the same magnitude as that of a present cosmological quantity, while it should be fixed by microphysics? 
This last question also opens the way to the famous coincidence problem associated to the cosmological constant.
\\
\\
Therefore, one could prefer suggesting a cosmological mechanism to justify the observed value of the cosmological constant.
Quintessence constitutes such an alternative explanation of DE 
that can be seen as a "\textit{varying}" cosmological constant \cite{peebles}.
The most common mathematical tool to model such an effect is the use of a real 
(electrically neutral— DE being dark) scalar field rolling down some self-interaction potential.
More precisely, the quintessence mechanism is implemented by the following theory:
\begin{eqnarray}
\label{quint}
S_{Quint}&=&\frac{1}{2\kappa_*}\int\sqrt{-g_*}d^4x\left\{R^*-2g_*^{\mu\nu}\partial_\mu\varphi\partial_\nu\varphi+V(\varphi)\right\}\nonumber\\
&&+S_m\left[\psi_m,g^*_{\mu\nu}\right],
\end{eqnarray}
with $\varphi$ the quintessence scalar field and $V(\varphi)$ its self-interaction potential. 
Scalar fields are often met in high-energy physics beyond the standard model like in supersymmetry 
(the only fundamental scalar in the standard model of elementary particles is the yet undiscovered Higgs boson). 
During a slow-roll phase of this massive scalar field 
(when its kinetic energy is much less than its self-interacting one), 
negative pressures are achieved, producing the desired cosmic acceleration if they violate the SEC.
Such a condition can be achieved when the scalar field freezes in a non-vanishing energy state due to
the damping provided by the cosmic expansion. 
Indeed, this damping arises because the scalar field energy density feeds the cosmic expansion all along cosmic history. 
The quintessence mechanism therefore relies strongly on an appropriate choice of self-interaction potential for the scalar field whose deep nature 
has still to be established. In particular, most of the potentials used in quintessence come out of an effective theory of high-energy physics, 
leaving this assumption a glimpse to new physics. These potentials can be quite sophisticated
and have to be well-shaped in order to reproduce Hubble diagram data or to exhibit interesting tracking properties \cite{steinhardt}.
Finally, one can say that the coincidence problem is somehow moved to the convenient choice of both an appropriate shape and
energy scale for
the self-interaction potential of an effective degree-of-freedom (the scalar field) representing a dynamical DE. Furthermore, 
deriving such quintessence models from high-energy physics in a self-consistent framework is a hard challenge, as
combining all the cosmological, gravitational and particle physics aspects raises many difficulties (see for instance \cite{brax}).
\\
\\
However, the quintessence scalar field minimally couples to gravitation and does not couple directly to ordinary matter.
In cosmology, this means that quintessence does only modify the background cosmic expansion.
But DE can also couple non-minimally to gravitation, if matter fields experience both DE and gravitation directly.
These new direct interactions between DE and ordinary matter yields a violation of the equivalence principle. 
Indeed, the gravitational binding energy therefore varies with the local intensity of these non-minimal coupling interactions.
The non-minimal coupling changes physical coupling constants or inertial masses and in consequence modifies the way energies have weighted in 
cosmic history.
In the case
where the scalar field directly affects the couplings to the gravitational field (the Einstein metric $g_{\mu\nu}$) 
and therefore
makes the gravitational coupling constant varying, we face the usual \textit{scalar-tensor} theories of gravitation:
\begin{eqnarray}
\label{st}
S_{TS}&=&\frac{1}{2\kappa_*}\int\sqrt{-g_*}d^4x\left\{R^*-2g_*^{\mu\nu}\partial_\mu\varphi\partial_\nu\varphi+V(\varphi)\right\}\nonumber\\
&&+S_m\left[\psi_m,A_m^2(\varphi)g^*_{\mu\nu}\right],
\end{eqnarray}
where $A_m(\varphi)$ denotes the constitutive coupling function to the Einstein metric (quintessence corresponds to $A_m(\varphi)=1$).
Those kinds of theories describe the
violation of the strong equivalence principle (SEP) (\cite{ts,convts}).
However, if we consider that the $70\%$ of missing energy is the whole contribution of such a non-minimally coupled
scalar field, then it might be difficult to match the present tests
of GR without assuming the non-minimal couplings to be \textit{extremely} weak (see for instance the so-called "extended quintessence" models
\cite{extq}). 
A possible solution to this problem is to consider a massive scalar field and
to render the effective mass
dependent on the background density (\cite{massd,chameleon}). 
In order to do so, a convenient self-interaction potential and a consequent violation of SEC have to be reintroduced.
Even worst, as soon as these couplings are non-vanishing, the likely endless domination of a non-minimally coupled scalar
field with quintessence potential will inescapably lead to a disastrous violation of the SEP in the future.
Therefore, if considering non-minimal couplings of DE to gravitation is a natural extension of quintessence, it nevertheless
rises many difficulties due to the stringent constraints on the equivalence principle and the constancy of fundamental constants. 
\\
\\
However, let us now put the DE mechanisms (\ref{eh}), (\ref{quint}) and (\ref{st}) in perspective.
In some sense, the simple cosmological constant can be seen as the tip of the iceberg of a deeper intriguing theory of gravitation.
In the framework of quintessence, it corresponds to the limiting case where the scalar field freezes in a non-vanishing energy
state. Quintessence itself can be seen as the limiting case of tensor-scalar gravity with negligible violation of SEP (negligible non-minimal
couplings). Finally, there is a generalization of the non-minimal couplings that embed the previous 
tensor-scalar theories: the case where the non-minimal couplings are not universal. This will constitutes the starting point
of the Abnormally Weighting Energy (AWE) Hypothesis \cite{fuzfa}.
\subsection{The AWEsome dust fluid dynamics}
In the AWE hypothesis, gravitation is no more ruled by the equivalence principle and
is therefore mediated by a pure spin 2 ($g_{\mu\nu}^*$) and a spin 0 ($\varphi$) degrees of freedom.
This last field rules the running of the gravitational couplings. Indeed, matter
couples to a metric which is screened by a conformal transformation written in terms of the scalar field $\varphi$. 
This screening of the metric $g_{\mu\nu}^*$ by $\varphi$ is a non-minimal and non-universal coupling:
it is different for the AWE than for ordinary matter.
This constitutes the violation of the weak equivalence principle (WEP).
More precisely, the energy
content of the universe is divided into three parts : a
gravitational sector with spin 2 and spin 0 components, a matter sector containing the usual
fluids of cosmology (baryons, photons, normally weighting dark matter if any, etc.)
and an AWE sector. The ordinary and abnormally weighting matter are assumed to interact exclusively through
their gravitational influence without any direct interaction.
The corresponding action can be written down in terms of the physical degrees
of freedom $g_{\mu\nu}^*$, $\varphi$, $\psi_m$ and $\psi_{awe}$ :
\begin{eqnarray}
\label{action1}
S&=&\frac{1}{2\kappa_*}\int\sqrt{-g_*}d^4x\left\{R^*-2g_*^{\mu\nu}\partial_\mu\varphi\partial_\nu\varphi\right\}\nonumber\\
&+&S_m\left[\psi_m,A_m^2(\varphi)g^*_{\mu\nu}\right]\\
&+&S_{awe}\left[\psi_{awe},A_{awe}^2(\varphi)g^*_{\mu\nu}\right],\nonumber
\end{eqnarray}
where $\kappa_*=8\pi G_*$ with $G_*$ is the "\textit{bare}" gravitational coupling
constant. In the previous action, $\varphi$ is a massless scalar field screening the metric $g^*_{\mu\nu}$, 
$S_{awe}$ is the action for the AWE
sector with fields $\psi_{awe}$ and $S_m$ is the usual matter
sector with matter fields $\psi_m$; $A_{awe}(\varphi)$ and $A_m(\varphi)$ being the constitutive coupling functions 
to the metric $g^*_{\mu\nu}$ for the AWE and matter sectors respectively.
The action (\ref{action1}) constitutes the so-called "\textit{Einstein frame}" 
of the physical, separated, degrees of freedom. In this frame, the metric
components are measured by using purely gravitational rods and
clocks, i.e. not build upon any of the matter fields nor the ones from the AWE sector. Therefore, this frame
does not correspond to a physically observable frame as we show in the next section. \\
\\
However, it is also important to notice that the AWE hypothesis naturally derives from more general effective theories
of gravitation motivated by string theory in which the couplings of the different matter fields to the dilaton are
not universal in general, and of order unity (see for instance \cite{damour2} and references therein).  This non-universality of the
gravitational couplings ($A_{awe}\ne A_m$) yields a violation of the
WEP: experiments using the new
AWE sector would provide a different inertial mass
than all other experiments. The reader should notice that action (\ref{action1}) of the
AWE hypothesis generalizes the models of DM-DE couplings. The case studied in \cite{pscora} corresponds
to $A_{awe}(\varphi)=\exp(\beta\varphi)$ and $A_m(\varphi)=1$ ($\psi_{awe}$ being DM fields), 
while chameleon cosmology \cite{chameleon} corresponds
to $A_{m,awe}(\varphi)=\exp(\beta_{m,awe}\varphi)$. 
\\
\\
Let us now turn on cosmology, and write down the field equations deriving from the action (\ref{action1}) upon the assumption of a flat Friedmann-Lemaitre-Robertson-Walker 
(FLRW)
universe with metric ($c=1$)
\begin{equation}
ds^2=-dt_*^2+a^2_*(t_*)dl_*^2, 
\label{flrw1}
\end{equation}
with $a_*(t_*)$ the scale factor
and $dl^2_*$ the Euclidean line element in the Einstein frame. We also consider the matter and AWE action as perfect fluids.
The Hamiltonian constraint of the action (\ref{action1}) gives
us the Friedmann equation:
\begin{equation}
\label{friedmann}
H_*^2=\left(\frac{\dot{a_*}}{a_*}\right)^2=\frac{\dot{\varphi}^2}{3}+\frac{\kappa}{3}\left(\rho^*_m+\rho^*_{awe}\right)
\end{equation}
where a dot denotes a derivation with respect to the time $t_*$ in the Einstein frame.
The acceleration equation can be written down:
\begin{equation}
\label{acc}
\frac{\ddot{a_*}}{a_*}=-\frac{2}{3}\dot{\varphi}^2-\frac{\kappa_*}{6}\left[\left(\rho^*_m+3p^*_m\right)+\left(\rho^*_{awe}+3p^*_{awe}\right)\right]\cdot
\end{equation}
Therefore, there cannot be any acceleration provided there is no violation of the
SEC in this frame if we assume pressureless fluids for usual matter and AWE. 
The expansion is then always decelerated $\ddot{a_*}<0$ in this frame and acceleration will occur only when we will move to the observable frame.
The Klein-Gordon equation ruling the scalar field dynamics is
\begin{eqnarray}
\label{kg_awe}
\ddot{\varphi}+3\frac{\dot{a_*}}{a_*}\dot{\varphi}&+&\frac{\kappa_*}{2}\alpha_m(\varphi)\left(\rho^*_m-3p^*_m\right)\nonumber\\
&+&\frac{\kappa_*}{2}\alpha_{awe}(\varphi)\left(\rho^*_{awe}-3p^*_{awe}\right)=0,
\end{eqnarray}
where $\alpha_i(\varphi)=\frac{d(\ln(A_i(\varphi))}{d\varphi}$ are the coupling functions. 
\\
\\
Let us now consider that the AWE sector is constituted by a pressureless fluid (``\textit{Abnormally Weighting Dark Matter}''; $p^*_{awe}=0$)
and focus on the matter-dominated era of the Universe ($p^*_m=0$). 
We can write down the conservation equations for the matter and the AWE sector separately (these sectors are decoupled according to
(\ref{action1})). For the matter sector, this equation reads $\nabla^*_\mu T^{\mu\; m,awe}_{*\nu}=\alpha_{m,awe} T_*^{m,awe} \partial_\nu \varphi$,
or, in terms of the FLRW ansatz (\ref{flrw1}):
\begin{eqnarray}
\nonumber
\dot{\rho}^*_{m,awe}+3\frac{\dot{a_*}}{a_*}\rho^*_{m,awe}=\alpha_{m,awe}(\varphi)\; \dot{\varphi}\rho^*_{m,awe}
\end{eqnarray}
which can be directly integrated to give
\begin{eqnarray}
\label{rhos}
\rho^*_{m,awe} &=& A_{m,awe}(\varphi) \frac{\mathcal{C}_{1,2}}{a_*^3}\cdot
\end{eqnarray}
In the above equation, $\mathcal{C}_{1,2}$ are arbitrary constants to be specified further. In the following,
we will denote by $R_i$ the ratio of $\mathcal{C}_1$ over $\mathcal{C}_2$: 
\begin{equation}
\label{Ri}
R_i=\mathcal{C}_1/\mathcal{C}_2\cdot
\end{equation}
It is possible then to rewrite equations (\ref{friedmann}), (\ref{acc}) and (\ref{kg_awe}) in a succinct form by using the quantity 
\begin{equation}
\label{rhot}
\rho_T=\rho_m+\rho_{awe}=\frac{\mathcal{A}(\varphi)\mathcal{C}_1}{a_*^3}
\end{equation}
where
\begin{equation}
\label{biga}
\mathcal{A}(\varphi)=A_m(\varphi)+\frac{A_{awe}(\varphi)}{R_i}
\end{equation}
is a resulting constitutive coupling function for the mixture of ordinary matter and AWE DM. 
Therefore, the violation
of the WEP by the AWE and matter dust fluids can be modeled by a tensor-scalar theory with only one dust fluid in which the constitutive coupling function $\mathcal{A}(\varphi)$
that results from the blend
is a linear combination of the constitutive coupling functions $A_m(\varphi)$ and $A_{awe}(\varphi)$ of the different sectors. 
\\
\\
Then, following the procedure initiated in \cite{convts} to reduce (\ref{kg_awe})
to an autonomous equation, we use the number of e-foldings : $\lambda=\ln (a_*/a^i_*)$ as a time variable (with $a^i_*$ is assumed here to mark the beginning of the
matter-dominated era, $a^i_*\approx 10^{-3}$). Using (\ref{friedmann}), (\ref{acc}), (\ref{rhot}), (\ref{biga}), 
the Klein-Gordon equation (\ref{kg_awe}) now reduces to (a prime denoting a derivative with respect to $\lambda$)
\begin{equation}
\label{kg2}
\frac{2\varphi''}{3-\varphi^{'2}}+\varphi'+\aleph(\varphi)=0
\end{equation}
where
\begin{equation}
\label{aleph}
\aleph(\varphi)=\frac{d\ln\mathcal{A}(\varphi)}{d\varphi}=\alpha_m(\varphi)+\frac{\alpha_{awe}(\varphi)-\alpha_m(\varphi)}{1+R_i \frac{A_m(\varphi)}{A_{awe}(\varphi)}}
\end{equation}
with
\begin{equation}
R_i \frac{A_m(\varphi)}{A_{awe}(\varphi)}=\frac{\rho^*_m}{\rho^*_{awe}}
\end{equation}
according to (\ref{rhos}).
Because of the vanishing pressures, equation (\ref{kg2}) is autonomous, as both dust fluids
have the same dependance on $a_*$ in their scaling laws (\ref{rhos}).
The last term in (\ref{kg2}) therefore contains all the information about the violation
of the WEP. Usual tensor-scalar theories in the matter-dominated era (see \cite{convts}) are easily retrieved if $\alpha_m(\varphi)=\alpha_{awe}(\varphi)=\aleph(\varphi)$ in
(\ref{aleph}),
which corresponds to no violation of the WEP and/or if the AWE fluid is sub-dominant ($\rho^*_m\gg \rho^*_{awe}$).
This last point is of first importance for the local tests of GR.\\
\\
The present tensor-scalar theory
exhibits a sophisticated convergence mechanism even in the case of very simple constitutive coupling functions $A_m(\varphi)$ and $A_{awe}(\varphi)$.
This feature is the key to reproduce DE effects in the AWE framework.
Indeed, let us study equation (\ref{kg2}) in terms of dynamical systems methods to charaterize the asymptotic dynamics
of the scalar field $\varphi$. 
This study is completely analogous of the case of usual tensor-scalar theories with WEP, but for the coupling function $\aleph(\varphi)=d\ln\mathcal{A}(\varphi)/d\varphi$.
The fixed points of the differential equation (\ref{kg2}) correspond to the extrema of $\mathcal{A}(\varphi)$ (zeros of $\aleph(\varphi)$)  and $\varphi'_\infty=0$ 
which
trivially corresponds to staying at rest at the extrema of the resulting coupling function $\mathcal{A}(\varphi)$ (\ref{biga}).
According to (\ref{biga}) and (\ref{aleph}), the fixed point $\varphi_\infty$ verifies $\aleph(\varphi_\infty)=0$, or more explicitely:
\begin{equation}
\label{phi_inf}
\alpha_m(\varphi_\infty)R_i\frac{A_m(\varphi_\infty)}{A_{awe}(\varphi_\infty)}+\alpha_{awe}(\varphi_\infty)=0\cdot
\end{equation}
For any set of coupling functions $A_m(\varphi)$, $A_{awe}(\varphi)$ for which (\ref{phi_inf}) admits a solution, the resulting coupling function
$\mathcal{A}$ has at least one extremum and there exists a finite value
of the effective gravitational coupling constant (denoted by $\tilde{G}_c(\varphi_\infty)$ as we shall see below) which is different from GR. 
This attracting value $\varphi_\infty$ depends
on the ratio of usual matter over abnormally weighting dust and is intermediate between the 
value of $\varphi$ for which $A_m(\varphi)$ is extremum (when $\rho^*_m\gg\rho^*_{awe}$) and the value of
$\varphi$ for which $A_{awe}(\varphi)$ is extremum (when $\rho^*_m\ll\rho^*_{awe}$).
\\
\\
\begin{figure}
\includegraphics[scale=0.5]{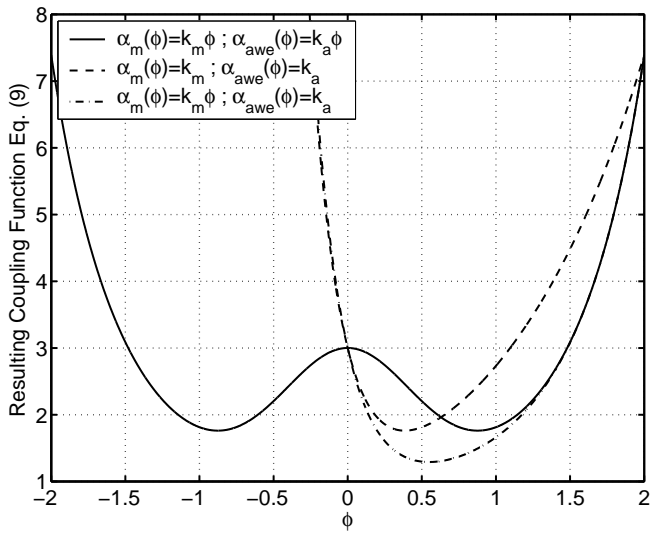}
\caption{Shapes of the resulting constitutive coupling function $\mathcal{A}(\varphi)$ (\ref{biga}) for three different sets of constitutive coupling functions
$A_m(\varphi)$ and $A_{awe}(\varphi)$
($R_i=0.5$ ; $k_m=1$ ; $k_{awe}=-5$) \label{fig1}}
\end{figure}
We can now illustrate that the attraction mechanism toward $\varphi_\infty$ exists for various choices of 
the couplings to the scalar field, even in the simplest cases. 
The shape of the resulting coupling function $\mathcal{A}(\varphi)$ (\ref{biga}) is presented in Figure \ref{fig1} 
where the attracting value $\varphi_\infty$ clearly appears as the mimimum of the resulting coupling. Models with $\ln(\mathcal{A}_{awe})\approx -\varphi^{2n}$ and
$\ln(\mathcal{A}_{m})\approx \varphi^{2n}$ ($n>0$) will exhibit a nice shape of double-well potential. The reader will find in the appendix several relations giving the position
of the attractor and the dynamics around it.\\
\\
The violation of the WEP strongly depends on the ratio of usual matter energy density over AWE. 
As both usual matter and AWE clusters, this violation is different with the scale considered. Furthermore, the AWE is assumed to be dark
(insensitive to the electromagnetic interaction for instance) and its gravitational collapse will be therefore quite different due to the absence
of the dissipative processes that allows usual matter like baryons to cluster so much compared to DM at small scales. 
This large abundance of usual matter at small scales therefore explains why the WEP is locally verified. 
Indeed, on small scales where AWE DM is sub-dominant $\rho^*_{awe}\ll\rho^*_m$, the value of the gravitational coupling is attracted toward $G_*$ 
($\alpha_m(\varphi_\infty)=0$). 
According to this argument, 
we will therefore conjecture that GR is verified
on the very small (sub-galactic) scales at which GR is well constrained by solar system or binary pulsar measurements. 
This will allow us in the following to explain the Hubble diagram of far-away supernovae only in terms of cosmic acceleration without
corrections due to the variation of the gravitational constant at small scales (see \cite{fuzfa} where this variation has been considered). 
Of course, the expected deviations from GR with the scale 
constitutes a key prediction of the AWE hypothesis but is left for future studies.
\\
\\
In the appendix, the reader will find a detailed analysis highlighting how the convergence mechanism toward GR is preserved despite the violation of WEP
brought by the non-universal couplings $A_m(\varphi)\ne A_{awe}(\varphi)$.
Before explaining how this revisited convergence mechanism can lead to cosmic acceleration, it is important to interprete the dynamics of the scalar field in
the appropriate way. This will be the subject of the next section.
\section{Cosmology with the AWE hypothesis}
The field equations we presented before were written in terms of the \textit{Einstein frame}, in which the physical degrees of freedom are separated.
In the context of the AWE hypothesis, we can build two other interesting frames or sets of variables using conformal transformations: 
the observable frame associated to ordinary matter and the AWE frame which can be seen as an observable frame for observers
made of AWE matter.\\
\\
Cosmology (and more generally everyday physics) is built upon observations based on usual matter (baryons, photons, etc.).
In the AWE hypothesis of action (\ref{action1}), 
this "\textit{normal}" matter couples universally to a unique metric, which we will denote $\tilde{g}_{\mu\nu}$, 
even though the coupling strength to this metric
varies in space-time. This metric $\tilde{g}_{\mu\nu}$ therefore defines the observable frame through the conformal transformation :
\begin{equation}
\label{gtilde}
\tilde{g}_{\mu\nu}=A_m^2\left(\varphi\right)g^*_{\mu\nu}\cdot
\end{equation}
However, this metric comprises scalar and tensorial degrees-of-freedom yielding different dynamics than the pure spin 2 Einstein metric
$g^*_{\mu\nu}$.
Under this transformation, the action of usual matter $S_m\left[\psi_m,A_m^2(\varphi)g^*_{\mu\nu}\right]$ writes down 
$\tilde{S}_m\left[\psi_m,\tilde{g}_{\mu\nu}\right]$ like in GR
and does not depend explicitly on $\varphi$. The energy-momentum conservation equations are $\tilde{\nabla}_\mu \tilde{T}^\mu_\nu=0$ with
$\tilde{\nabla}$ the covariant derivative associated to the effective metric $\tilde{g}_{\mu\nu}$. Therefore, the components of the stress-energy tensor
in the observable frame will have the same scaling law than in GR.
The relations between the components of the stress-energy tensor in the observable and Einstein frames are given by
\begin{equation}
\tilde{T}^{m\;\mu}_\nu=A^{-4}_{m}(\varphi)T^{m\;\mu}_{*\nu}\cdot
\end{equation}
\\
\\
At this point, we would like to point out that there exists another interesting frame, the AWE frame, defined as the one into
which all the abnormally weighting fields $\psi_{awe}$ couple universally to the effective metric:
\begin{equation}
\label{barg}
\bar{g}_{\mu\nu}=A_{awe}^2(\varphi)g^*_{\mu\nu}\cdot
\end{equation}
As there is no direct coupling between matter and AWE fields (no dependency of the matter action upon $\psi_{awe}$ and vice-versa), the
same arguments for the stress-energy conservation as above holds for this AWE frame. In particular, we now have
$$
\bar{T}^{awe\;\mu}_\nu=A^{-4}_{awe}(\varphi)T^{awe\;\mu}_{*\nu},
$$
with $T^{awe\;\mu}_{*\nu}$ is the stress-energy tensor of the AWE sector in the Einstein frame and
where $\bar{T}^{awe\;\mu}_\nu$ behaves like in GR.
The observable components of the AWE stress-energy tensor write down
\begin{equation}
\tilde{T}^{awe\;\mu}_\nu=A^{-4}_{m}(\varphi)A^{4}_{awe}(\varphi)\bar{T}^\mu_{awe\;\nu}
\end{equation}
The AWE hypothesis therefore lead us to a gravitational theory in which there are two effective metrics, $\tilde{g}_{\mu\nu}$ and 
$\bar{g}_{\mu\nu}$
which couple \textit{universally} to a separate set of physical fields $\psi_m$ and $\psi_{awe}$. 
The absence of a unique effective metric is a consequence of the
violation of the WEP.
But these two metrics are simply related by a conformal transformation 
\begin{equation}
\tilde{g}_{\mu\nu}=\left(\frac{A_m(\varphi)}{A_{awe}(\varphi)}\right)^2\bar{g}_{\mu\nu}
\end{equation}
so that there is only one set of underlying tensorial degrees-of-freedom: the $g_{\mu\nu}^*$'s. There is only one space-time with a curvature felt differently by matter
and AWE. Therefore, 
the violation of the WEP considered here is somehow minimal in the sense that the theory of gravitation given by (\ref{action1}) simply has two different
gravitational running coupling constants: 
\begin{equation}
\label{gc}
\tilde{G}_c=A_m^2(\varphi)G_*
\end{equation}
for the usual matter sector and 
$$
\bar{G}_c=A_{awe}^2(\varphi)G_*
$$
for
the AWE sector.
\\
\\
Let us now move on the observable quantities of the AWE cosmology. We begin by assuming
a FLRW parametrization of the observable metric $\tilde{g}_{\mu\nu}$:
\begin{equation}
\tilde{ds}^2=-d\tilde{t}^2+\tilde{a}^2(\tilde{t})d\tilde{l}^2, 
\end{equation}
where the observable scale factor is given by
\begin{equation}
\label{atilde}
\tilde{a}(\tilde{t})=A_m(\varphi)a_*(t_*)=\left(\frac{\tilde{G}_c}{G_*}\right)^{1/2}a_*(t_*)
\end{equation}
with $\tilde{G}_c$ the effective gravitational coupling constant (\ref{gc}) for the matter sector and the element of observable synchronous time reads
\begin{equation}
d\tilde{t}=A_m(\varphi)dt_*
\end{equation}
according to (\ref{gtilde}). 
Performing the conformal transformation
(\ref{gtilde}) on the field equations in the Einstein frame (\ref{friedmann}) (see also \cite{fuzfa}), we get
for the observable expansion rate
\begin{eqnarray}
\tilde{H}^2=&&\frac{8\pi\tilde{G}_c}{3}\left(\tilde{\rho}_m+\tilde{\rho}_{awe}\right)\times\nonumber\\
&&\left(1+\frac{\varphi^{'2}\left(1+3\alpha_m^2\right)+6\alpha_m\varphi'}{3-\varphi^{'2}}\right)
\label{hobs}
\end{eqnarray}
with $\tilde{\rho}_{m,awe}=A^{-4}_m\rho^*_{m,awe}$ and
\begin{equation}
\tilde{H}=\frac{1}{\tilde{a}}\frac{d\tilde{a}}{d\tilde{t}}=A^{-1}_m(\varphi)H_*(1+\alpha_m(\varphi)\varphi')\cdot
\label{htilde}
\end{equation}
This allows us to write down the following density parameters for matter and abnormally weighting dust
\begin{eqnarray}
\tilde{\Omega}_{m,awe}&=&\frac{8\pi \tilde{G}_c \tilde{\rho}_{m,awe}}{3\tilde{H}^2}
\label{omegas1}
\end{eqnarray}
and an effective density parameter for the scalar field:
\begin{equation}
\tilde{\Omega}_{\varphi}=\left(\tilde{\Omega}_{m}+\tilde{\Omega}_{awe}\right)\frac{\varphi'\left(1+3\alpha_m^2\right)+6\alpha_m}{3-\varphi^{'2}}\varphi'\cdot
\label{omegas2}
\end{equation}
with the constraint $\tilde{\Omega}_{m}+\tilde{\Omega}_{awe}+\tilde{\Omega}_{\varphi}=1$ (flat universe). Before going any further, it should be noted that 
the ratio of the matter and AWE density parameters is given by 
\begin{equation}
\frac{\tilde{\Omega}_m}{\tilde{\Omega}_{awe}}=R_i \frac{A_m(\varphi)}{A_{awe}(\varphi)},
\label{rhom_rhoa}
\end{equation}
so that $R_i$ represents this ratio when $A_m(\varphi)=A_{awe}(\varphi)$. $R_i$ will be one of the free parameter of the present DE model.
\\
\\
The acceleration equation in the observable frame is given by:
\begin{eqnarray}
\frac{1}{\tilde{a}}\frac{d^2\tilde{a}}{d\tilde{t}^2}&=&
-\frac{4\pi \tilde{G}_c}{3}\left(\tilde{\rho}_m+\tilde{\rho}_{awe}\right)\times\nonumber\\
&&\left(1-\frac{2\varphi'}{3-\varphi^{'2}}\left(\varphi'\left(\frac{d\alpha_m}{d\varphi}-\frac{2}{3}\right)-2\alpha_m\right)\right)\nonumber\\
&&-4\pi \tilde{G}_c\alpha_m\left(\alpha_m\tilde{\rho}_m+\alpha_{awe}\tilde{\rho}_{awe}\right)
\label{obs_acc}
\end{eqnarray}
Although the energy densities are always positive and therefore will provide only a deceleration of the cosmic expansion,
their couplings to the scalar field can lead to cosmic acceleration under particular circumstances (this might occur for instance
if $\alpha_{awe}<0$ and $\rho^*_{awe}\gg \rho^*_m$). 
Cosmic acceleration in the observable frame is 
provided by $A_m(\varphi)$ (i.e., the coupling constant $\tilde{G}_c$) and not $a_*$ in (\ref{atilde}) which is related to an
acceleration of the scalar field itself resulting from the competition between usual matter (through $A_m(\varphi)$) and AWE (through $A_{awe}(\varphi)$) in (\ref{kg2}).\\
\\
We can now search for a translation of the AWE cosmology in usual FLRW cosmology.
In order to do so, 
one can search for an effective DE density $\tilde{\rho}_{DE}$ together with its effective equation of state by matching the terms in (\ref{hobs}) and (\ref{obs_acc})
to:
\begin{eqnarray}
\tilde{H}^2&=&\frac{8\pi\tilde{G}_c}{3}\left(\tilde{\rho}_{m,T}+\tilde{\rho}_{DE}\right)\\
\frac{1}{\tilde{a}}\frac{d^2\tilde{a}}{d\tilde{t}^2}&=&-\frac{4\pi \tilde{G}_c}{3}\tilde{\rho}_{m,T}-\frac{4\pi \tilde{G}_c}{3}\tilde{\rho}_{DE}(1+3\omega_{eff})
\end{eqnarray}
where $\tilde{\rho}_{m,T}=\tilde{\rho}_{m}(1+R_i^{-1})$ is the total amount of ordinary matter and DM\footnote{
The observable energy density of AWE can be written down: 
$$
\tilde{\rho}_{awe}=\left(\frac{A_{awe}(\varphi)}{A_{m}(\varphi)}-1\right)\frac{\tilde{\rho}_m}{R_i}+\frac{\tilde{\rho}_m}{R_i}
$$
with (\ref{Ri}).} 
rescaling exactly in $\tilde{a}^{-3}$.
After some manipulations, we find the following effective equation of state:
\begin{eqnarray}
\label{weff}
\omega_{eff}&=&\left\{-\left(\tilde{\Omega}_{m}+\tilde{\Omega}_{awe}\right)\times\right.\nonumber\\
&&\left[\varphi'^2\left(2\frac{d\alpha_m}{d\varphi}-\frac{1}{3}+3\alpha_m^2\right)+2\alpha_m\varphi'\right]\nonumber\\
&& \left.+3\alpha_m\left(3-\varphi^{'2}\right)\left(\alpha_m\tilde{\Omega}_m+\alpha_{awe}\tilde{\Omega}_{awe}\right)\right\}\times\nonumber\\
&&\left\{3\left(3-\varphi^{'2}\right)\left(1-\tilde{\Omega}_m(1+R_i^{-1})\right)\right\}^{-1}
\end{eqnarray}
This equation of state asymptotically vanish at the attractor $\varphi_{\infty}$. We will also
see that phantom dark energies $\omega_{eff}<-1$ can easily be achieved, even today at the opposite of the results in \cite{pscora}.\\
\\
The existence of the attractor (\ref{phi_inf}) implies that the density parameters (\ref{omegas1}) freeze
at a constant ratio, when $a_*\rightarrow \infty$:
\begin{equation}
\label{rinf}
\frac{\tilde{\Omega}_m(\varphi_\infty)}{\tilde{\Omega}_{awe}(\varphi_\infty)}=R_i \frac{A_m(\varphi_\infty)}{A_{awe}(\varphi_\infty)}=R_{\infty}
\end{equation}
which defines another free parameter of the AWE DM model, $R_\infty$. Furthermore, as this fixed point is stable,
the scalar field kinetic energy asymptotically vanishes, leaving the other density parameters freezing at constant values given by
\begin{eqnarray}
\tilde{\Omega}_m(\varphi_\infty)&=&\frac{R_\infty}{1+R_\infty}\nonumber\\
\tilde{\Omega}_{awe}(\varphi_\infty)&=&\frac{1}{1+R_\infty}
\end{eqnarray}
($\tilde{\Omega}_\varphi(\varphi_\infty)=0=1-\tilde{\Omega}_m(\varphi_\infty)-\tilde{\Omega}_{awe}(\varphi_\infty)$).
\section{Dark Energy as the anomalous gravity of Dark Matter}
We can now illustrate how the AWE DM fluid explains the accelerated expansion measured by the far-away supernovae and how
its anomalous gravity plays the role of DE.
\subsection{Building a specific model}
Let us first choose a set of constitutive coupling functions\footnote{The reader will find other parametrizations in the appendix.} 
$A_m(\varphi)$ and $A_{awe}(\varphi)$:
\begin{eqnarray}
\left\{
\begin{tabular}{l}
$A_m(\varphi)=\exp\left(k_m\frac{\varphi^2}{2}\right)$\\
\\
$A_{awe}(\varphi)=\exp\left(k_{awe}\frac{\varphi^2}{2}\right)$\\
\end{tabular}
\right.\cdot
\label{model3}
\end{eqnarray}
With such a parametrization, the attraction mechanism guarantees an asymptotic solution which looks like GR, but with a different value of the gravitational coupling 
on cosmological scales. For the sake of simplicity, we therefore consider  that the scalar field starts moving at the beginning of the matter-dominated era 
with $(\varphi, \varphi')\approx (0,0)$ at $a_i^*\approx 10^{-3}$, which corresponds to start moving away from GR with the value of the gravitational coupling
corresponding to ordinary matter at the end of the radiative era. We refer the reader to section IV C. for a preliminary discussion on coincidence and
sensitivity to the initial conditions.\\
\\
The number of free parameters of the AWE DM mechanism in the present parametrization is three. Indeed, it is required to know
the relative amount of energy in the AWE and matter sectors  at start $R_i=\rho_m^*/\rho_{awe}^*(\varphi_i=0)$ 
and the free parameters of the coupling functions $A_m(\varphi)$ and $A_{awe}(\varphi)$ to the scalar field. 
For instance, the model (\ref{model3}) have three free parameters : $R_i$, $k_m$ and $k_{awe}$. 
Once these parameters are given, the ratio of matter over AWE $R_{\infty}$ to which their energy densities finally freeze is determined by 
the position of the attractor $\varphi_\infty$ (\ref{phi_inf}). Then, the integration of (\ref{kg2}) (or the full system (\ref{friedmann}), (\ref{acc}) and (\ref{kg_awe}))
leads to the observable density parameters today $\tilde{\Omega}_i(\tilde{z}=0)$. This is why we will rather characterize the models in terms of
the parameters $(R_i,\;\tilde{\Omega}_m(\tilde{z}=0)=\tilde{\Omega}_m^0,\;R_\infty)$ as the first two can be measured independently of the Hubble diagram of type Ia supernovae.
\subsection{Cosmological Discussion}
\begin{figure}
\includegraphics[scale=0.4]{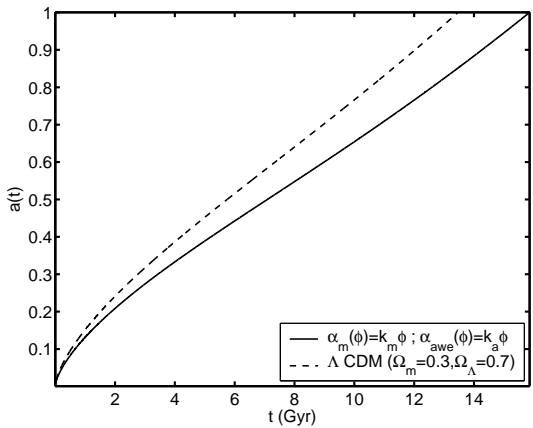}
\caption{Cosmic expansion $\tilde{a}(\tilde{t})$ for the model (\ref{model3}) (see also Table \ref{tab1}) and comparison with the concordance model
$\Lambda CDM$ (for which the $\chi^2$ per $dof$ for the same data set is $1.03$) \label{fig2}}
\end{figure}

\begin{figure}
\includegraphics[scale=0.4]{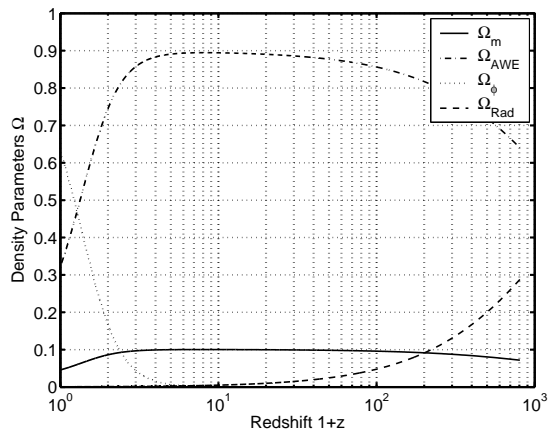}
\caption{Evolution of the density parameters $\tilde{\Omega}_i$ with the redshift $1+\tilde{z}$ 
for model (\ref{model3}) \label{fig3}}
\end{figure}

\begin{figure}
\includegraphics[scale=0.4]{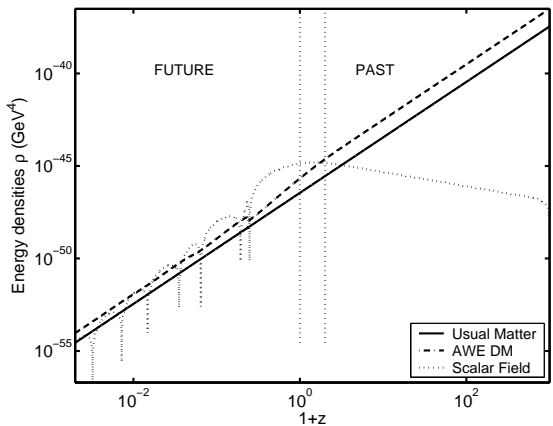}
\caption{Evolution of the energy densities $\tilde{\rho}_i$ with the redshift $1+\tilde{z}$ 
for model (\ref{model3}) (vertical lines represent the redshift range of the data set used for calibration)\label{fig4}}
\end{figure}

\begin{figure}
\includegraphics[scale=0.4]{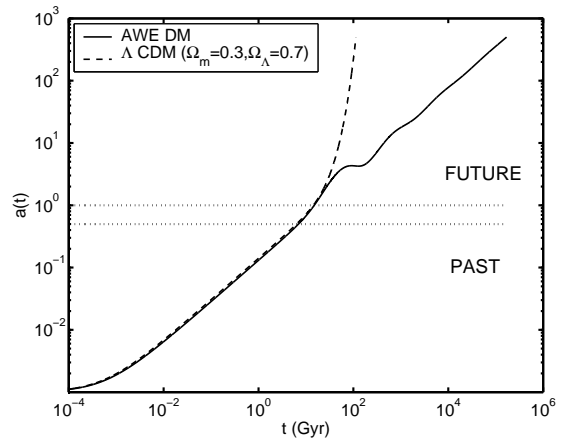}
\caption{Fate of the universe in the concordance model $\Lambda CDM$ and in the AWE DM model (\ref{model3}) (see also Table \ref{tab1}).
\label{fig5}}
\end{figure}

\begin{figure}
\includegraphics[scale=0.4]{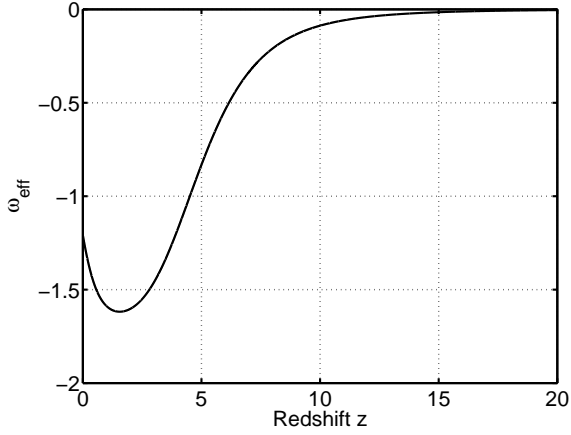}
\caption{Evolution of the effective equation of state $\omega_{eff}$ (\ref{weff}) with the redshift $\tilde{z}$ for model (\ref{model3}) \label{fig6}}
\end{figure}

\begin{figure}
\includegraphics[scale=0.4]{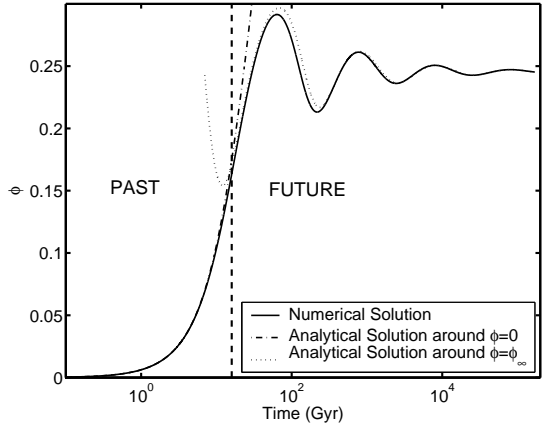}%
\caption{Evolution of the scalar field for model (\ref{model3}) and comparison with analytical approximations near the fixed points.
The vertical line represents the present state of the universe
\label{fig7}}
\end{figure}

Let us now calibrate these free parameters with current available data on the type Ia supernovae Hubble diagram \cite{snls}.
We show in the appendix how this Hubble diagram can be retrieved from the most simple shapes
of coupling functions $\alpha_i(\varphi)$, therefore illustrating how cosmic acceleration is a generic prediction of the AWE DM mechanism. 
According to what we have seen in section II, we can expect that GR is verified on small scales, meaning that
type Ia supernovae can be considered as standard candles. Therefore, their Hubble diagram can only be explained in terms of cosmic acceleration
without any correction
from a possible variation of Newton's constant (see \cite{fuzfa} for an AWE model without this prescription).
The distance moduli of type Ia supernovae are given by:
\begin{equation}
\label{muz} \mu(\tilde{z})=m-M=25+5\ln_{10} d_L(\tilde{z}),
\end{equation}
where $d_L(\tilde{z})$ is the luminous distance
(in Mpc) given by $d_L(\tilde{z})=(1+\tilde{z})\tilde{H}_0\int_0^{\tilde{z}}
d\tilde{z}/\tilde{H}(\tilde{z})$
for a flat universe ($\tilde{H}_0$ is the observed value of the
Hubble constant today). The expansion rate $\tilde{H}(\tilde{z})$
has to be estimated in the observable frame related to usual matter
(\ref{htilde}) as
indicated in section III. 
\\
\\
Table \ref{tab1} gives the calibrated values of the free parameters $R_i$, $R_\infty$ and $\tilde{\Omega}_m^0$ for various models
(see the appendix for the corresponding parametrizations) as well as other interesting related
quantities like the value of the density parameter for AWE and the age of the universe $t_0$. We remind
the reader that $\tilde{\Omega}_\varphi$ can be either negative or positive (see (\ref{omegas2})).
Very simple parametrizations of the constitutive coupling functions $A_m(\varphi)$ and $A_{awe}(\varphi)$ allow to account fairly and almost equivalently with data despite their
different definitions. This illustrates the robustness of the AWE hypothesis applied to dark matter.
\begin{table}
\begin{ruledtabular}
\begin{tabular}{|c|c|c|c|c|c|c|c|}
Model & $R_i$ & $R_\infty$ &$\tilde{\Omega}_m^0$ & $\tilde{\Omega}_{awe}^0$ & $t_0 (Gyr)$ & $\chi^2/dof$\\
(\ref{model3}) & $0.11$ & $0.31$ & $0.04$ &$0.26$ & $15.9$ & $1.03$\\
(\ref{model1}) & $4.5\times 10^{-3}$ & $9.7$ & $0.73$ &$0.12$ & $10.6$ & $1.05$\\
(\ref{model2}) & $2.5\times 10^{-3}$ & $1.8$ & $0.62$ &$0.43$ & $9.8$ & $1.03$\\
\end{tabular}
\end{ruledtabular}
\caption{Cosmological parameters for different parametrizations of the AWE DM model, as calibrated by Hubble diagram data \cite{snls}
\label{tab1}}
\end{table}
However, one should be careful while considering the parameters in table \ref{tab1} 
in the light of other cosmological tests as these last should also be performed
by taking properly into account the anomalous gravity of dark matter. Our aim here is to show that the AWE DM model can account easily for the observed cosmic acceleration
without requiring
well-shaped sophisticated functions. Figure \ref{fig2} gives the evolution of the scale factor (\ref{atilde}) for
the model (\ref{model3}), with a value of $\tilde{H}_0=70 km/s/Mpc$. 
Also illustrated is the concordance $\Lambda CDM$ model ($(\Omega_m,\Omega_\Lambda)=(0.3, 0.7)$). 
In model (\ref{model3}), the cosmic acceleration is achieved during the accelerated deviation of $\varphi$ from
the unstable point at $\varphi=0$.
\\
\\
In Figure \ref{fig3}, the evolution of the density parameters (\ref{omegas1}) and (\ref{omegas2}) of
ordinary matter, AWE DM and scalar field is given.
The AWE DM is dominating the energy content of the universe until $\tilde{a}\approx 0.68$ where its density parameter is joined by the one
of the rising scalar field. This epoch constitutes the coincidence between the AWE DM and the DE it produces through the scalar field $\varphi$.
The Universe has been dominated by AWE DM since the end of the radiative era and the violation of the WEP by the AWE DM has induced
a violation of the SEP on large-scales for the ordinary matter sector.
With the consequent growth of its gravitational coupling, the ordinary matter has experienced an increasingly stronger cosmic expansion.
We have in this scenario the following density parameters today: $\tilde{\Omega}_m^0=0.04$,
and $\tilde{\Omega}_{AWE}^0=0.26$. However, it should be noted that, up to now, we did not proceed with any assumption about the exact content of the ordinary matter sector, except that it 
comprises at least baryons on which the WEP is well tested. This ordinary matter sector might well include some amount of \textit{normally weighting} dark matter in addition to the baryons. 
Addressing the question whether one could identify or not the matter and AWE sectors to baryons and cold dark matter respectively requires to estimate the values of $\tilde{\Omega}_m^0$, $\tilde{\Omega}_{awe}^0$ with cosmological tests like the Hubble diagram method for instance. According to Table 1, the answer depends on the shape of the matter and AWE constitutive coupling functions $A_m(\varphi)$ and $A_{awe}(\varphi)$. In the model (\ref{model3}), 
one could therefore well identify ordinary matter with baryons and AWE with dark matter. This would give an elegant explanation of the various energy components of the concordance model and the deeper physical relations they maintain.
\\
\\
In Figure \ref{fig4}, the past and future evolutions of energy densities of the various components are represented\footnote{The absolute value of the scalar field energy
density $\tilde{\rho}_\varphi$, which can be negative according to (\ref{omegas2}), has been represented.}.
The scalar field will be dominant until we reach the attractor $\varphi_\infty$ at $\tilde{z}=-0.7$ (in approximately $50$ billion years,
see also Figures \ref{fig5} and \ref{fig7}) and $\varphi$ starts oscillating around $\varphi_\infty$. These oscillations will appear in the future
cosmic expansion, as can be seen in Figure \ref{fig5} and will take thousands and thousands of billion years to be damped to a negligible value.
More precisely, the amplitude of the energy density of the scalar field becomes sub-dominant after $\tilde{a}\approx 10^{2}$ (Figure \ref{fig4}),
at $\tilde{t}\approx 10^4 Gyr$. The final state of the universe, once the oscillations of the scalar field become negligible, is constituted by an Einstein-de Sitter 
expansion phase.
Indeed, the scalar field slowly freezes ($\varphi'\rightarrow 0$ and $\tilde{\Omega}_\varphi\rightarrow 0$) 
near the attractor $\varphi_\infty$, so that the matter and AWE DM energy densities will redshift as $\tilde{a}^{-3}$ 
(see Figure \ref{fig4}). At this distant epoch, gravitation will be described by a theory with purely tensorial degrees-of-freedom (the scalar sector being frozen) like GR but
with $\tilde{G}_c(\varphi_\infty)$ as a coupling constant instead of the bare coupling $G_*$. 
During the competition between AWE and ordinary matter for ruling the equivalence principle, 
cosmic acceleration and deceleration phases have been achieved several times before the final stabilization of the large scale gravitational
coupling constant $\tilde{G_c}$. 
Therefore, GR-like behavior appears both at the beginning (since $\tilde{z}\approx 1000$ the scalar field slowly rises) and at the end of the mechanism
($\tilde{z}= -1$). The deviation from standard FLRW cosmic expansion due to the induced violation of SEP on ordinary matter
disappears once a new equilibrium has been reached
for the gravitational coupling constant. In Figure \ref{fig5}, the concordance model $\Lambda CDM$ has been represented in order to illustrate the different
visions of the fate of the Universe. The cosmological constant yields an endless cosmic expansion which becomes exponential (de Sitter regime) in a few dozens billion years
(Figure \ref{fig5} near $\tilde{a}\approx 10$). The AWE DM model predicts also an eternal expansion but not an endless cosmic acceleration.
\\
\\
Let us give a characterization of the cosmic expansion in the AWE DM model in terms of an effective equation of state in a standard FLRW
framework (see section III).
We remind that the present DE mechanism is essentially based on a competition between different non-minimal couplings
and not on a violation of the strong energy condition (which is never violated here in the Einstein frame as the fluids are
pressureless). The present effective equation of state is for non-minimally coupled models and contains the information 
on both the behavior of the scalar field and the AWE DM fluid (see (\ref{weff})).
One can therefore expect a very different behavior for the equation of state 
$\omega_{eff}$ than for standard FLRW cosmology. This is illustrated in Figure \ref{fig6}. The equation of state starts
moving from $0$ (dust-dominated universe) after $\tilde{z}\approx 20$ to cross the equation of state of a cosmological constant ($\omega_{\Lambda}=-1$) near $\tilde{z}\approx 5$, then reverse at $\tilde{z}\approx 2$ to finally reach $\omega_{eff}=-1.2$ today. This "\textit{ghost-like}" equation of state is responsible for a stronger cosmic acceleration and the larger
age of the universe found in model (\ref{model3}). In the future, the equation of state will enter a damped oscillating regime, driven by the underlying
evolution of the scalar field. According to what we have seen above for the fate of the Universe, $\omega_{eff}$ eventually vanishes and
the universe recovers a dust-dominated expansion regime.
\\
\\
\subsection{About the coincidence problem}
Figure \ref{fig7} illustrates the validity of the analytical approximations given in the appendix ((\ref{phi_osc}) and (\ref{phi_mon})) around the fixed points $\varphi=\varphi_\infty$ 
and $\varphi=0$ of model (\ref{model3}).
These expressions will allow us to discuss the coincidence problem for this DE mechanism in the framework of model (\ref{model3}).  
The cosmic coincidence is reached when $\tilde{\Omega}_\varphi\approx \tilde{\Omega}_m +\tilde{\Omega}_{awe}$.
Using (\ref{omegas2}) in the non-relativistic limit ($\varphi^{'2}\ll 3$) and (\ref{phi_mon}), we find that the observable scale factor of coincidence $\tilde{a}_c$
is given by
\begin{eqnarray}
\tilde{a}_c&=&a_*^i\exp\left(\frac{2\left(\sqrt{2}-1\right)}{3\left[\sqrt{1-\frac{8}{3}K_0}-1\right]}\right)\times\\
&&\left(\frac{2(\sqrt{2}-1)\left(1-\frac{8}{3}K_0\right)}{\aleph(\varphi_i)\alpha_m(\varphi_i)\left(\sqrt{1-\frac{8}{3}K_0}+1\right)}\right)^{2/\left[3(\sqrt{1-\frac{8}{3}K_0}-1)\right]}\nonumber
\label{ac}
\end{eqnarray}
where we set
\begin{equation}
\frac{d\aleph(\varphi)}{d\varphi}|_{\varphi\approx\varphi_i}=K_0
\end{equation}
(see also the appendix). Eq. (\ref{ac}) agrees with numerically computed values by a few percent accuracy.
\begin{figure}
\includegraphics[scale=0.4]{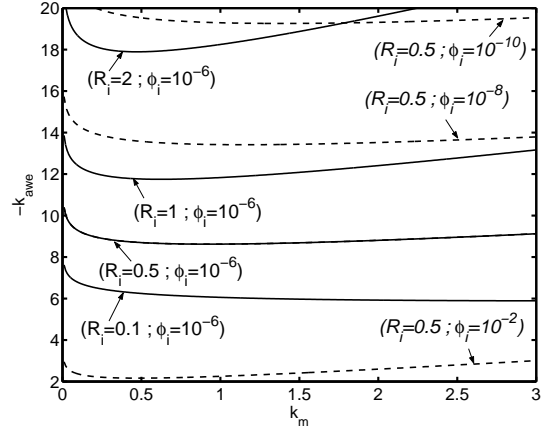}%
\caption{Values of free parameters $R_i$, $k_m$, $k_{awe}$ and initial condition $\varphi_i$ for which the coincidence occur at $\tilde{a}=0.5$ in model (\ref{model3})
($a_*^i=10^{-3}$)
\label{fig8}}
\end{figure}
Figure \ref{fig8} represents the values of the parameters $k_m$ and $k_{awe}$, given initial conditions $R_i$ and $\varphi_i$ for which the coincidence
occurs at $\tilde{a}_c=0.5$. There is a wide range of values of the coupling constants $k_m$ and $k_{awe}$ for which cosmic coincidence is explained,
even starting from very different initial conditions in the distribution of AWE and ordinary matter $R_i$, or in the initial value of the scalar field $\varphi_i$. Furthermore,
the magnitude of $k_m$ and $k_{awe}$ required to justify the coincidence is of order unity which means that the amplitude of the coupling
to the scalar field is of the same order of magnitude than the coupling $\kappa_*=8\pi$ (in units of $G_*$) to the metric tensor $g^*_{\mu\nu}$. The cosmic coincidence appears
as a consequence of
this very natural assumption in non-minimally coupled tensor-scalar theories of gravitation. 
For what concerns the tuning of the free parameters, these last do not need to vary by several orders of magnitude
if one changes slightly the time of coincidence, at the opposite of the cosmological constant. The amplitude of $k_m$ and $k_{awe}$ of order unity ensure a typical arising, duration and end of cosmic acceleration during matter-dominated era
as the AWE DM mechanism could not have occured during
the radiative era (radiation is ruled  by WEP). Therefore, the AWE DM model improves the coincidence and fine-tuning
problems of the cosmological constant. 
\subsection{AWE Dark Matter as a time-dependent inertial mass}
\begin{figure}
\includegraphics[scale=0.4]{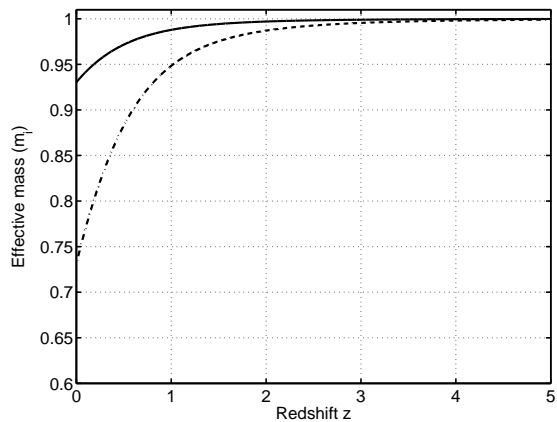}%
\caption{Evolution of the effective masses $m_*$ (straight line) and $\tilde{m}$ (dash-dotted line) (in units of the mass at $\varphi=0$) 
with the redshift $\tilde{z}$ for model (\ref{model3}) \label{fig9}}
\end{figure}

Let us now interprete this AWE mechanism in terms of a variation of the inertial mass of the dark matter particles.
This can be done by assuming that the AWE DM fluid is made of a collection of non-relativistic particles in such a way
that its energy density can be rewritten as $\rho=m\times n/a^3$ with $n$ the numerical density of particles per unit volume (assumed conserved)
and $m$ the inertial mass of the particles. Doing so, we can define two types of effective masses for AWE DM, the first is
for the Einstein frame $m_*$ (computed from $\rho_{awe}^*$):
\begin{equation}
m_*(\varphi)=A_{awe}(\varphi)
\end{equation}
and the second is the observable effective mass (computed from $\tilde{\rho}_{awe}$):
\begin{equation}
\tilde{m}(\varphi)=\frac{A_{awe}(\varphi)}{A_{m}(\varphi)}\cdot
\end{equation}
This last expression is also proportional to the ratio of the AWE DM and ordinary matter
energy densities (or equivalently density parameters) in any frame (see (\ref{rhom_rhoa}) and
other related relations in section III). The variation of these inertial masses for the calibrated model (\ref{model3}) is given in Figure \ref{fig9}.
The cosmic acceleration measured by the supernovae needs a variation of $\tilde{m}$ of about $30\%$ in the redshift range $\tilde{z}=0$ to $5$.
The inertial mass $\tilde{m}$ of the AWE DM particles starts varying at $\tilde{z}\approx 4$ and finally freezes 
at $\tilde{m}=0.38$ ($m_*=0.8$) at $\tilde{z}=-1$ when $\varphi$ reaches the attractor. This variation
of the inertial mass of AWE DM violates the WEP which yields an increasingly stronger gravitational coupling for
ordinary matter appearing as a cosmic acceleration.
\section{Conclusion}
The theoretical explanation of dark energy is probably one the most challenging and promising issue for both cosmology and fundamental physics. The long-time discarded and often controversed cosmological constant is one the main theoretical pillar of the concordance model in cosmology, together with the puzzling nature of dark matter.
Cosmological mechanisms of DE, like quintessence and non-minimal couplings, that aims to go beyond the simple cosmological constant to cure its intricate difficulties, make
this constant appear as the tip of the iceberg of an underlying more fundamental theory of gravitation.
Most of such theories that go beyond GR suggest a violation of the weak and/or strong equivalence principles. The AWE hypothesis, suggested in
\cite{fuzfa}, consists of applying this suggestion to the DE problem by assuming that this energy abnormally weights.
In terms of tensor-scalar theory of gravitation, there are two distinct matter sectors (the abnormally weighting DE 
and ordinary matter) that only interact through gravitational interactions and 
couples differently to the tensorial gravitational degrees of freedom. This is a minimum violation of the WEP, that yield
temporary deviations from GR resulting in cosmic acceleration (and possible variation of $G$, see also \cite{fuzfa}).
The anomalous gravity of DE implies a violation of the strong equivalence principle for the gravitational interactions. 
As a consequence, the varying gravitational coupling modifies the strength of the background cosmic expansion experienced by usual matter
so that it becomes possible not only to explain DE effects but also to consider a unified approach to DE and DM. \\
\\
In this paper, we have applied this idea to a pressureless fluid called the abnormally weighting dark matter (AWE DM). More precisely, we have considered that both 
the AWE and ordinary matter sectors
are made of pressureless fluids with different couplings to gravitation. We have shown that the convergence mechanism toward GR might be conserved, despite the violation of WEP, provided very general
conditions. The violation of the WEP therefore appears as an auxilliary driving force competing with the coupling to ordinary matter. This new driving force depends on
the ratio of ordinary matter over abnormally weighting one in such a way that the weak equivalence principle can be locally restored
when the last is sub-dominant. As the gravitational collapse is expected to be very different for the two matter sectors
(ordinary matter clusters much more as a consequence of the various dissipative processes it undergoes), this opens the possibility of
retrieving locally the precision of current tests of GR (a similar explanation has been invoked in previous works \cite{massd,chameleon}). 
\\
\\
However, on cosmological scales where the AWE DM is dominant, gravitation
is no longer described neither by GR nor by usual tensor-scalar theories with WEP.  The final state of cosmological evolution is indeed 
described by a GR-like theory for which the value of the
gravitational coupling constant is different than the bare one of GR and depends on the properties of AWE DM.
We have paid a particular attention of
re-writting the whole dynamics in terms of the effective metric to which ordinary matter universally couples. This definition of
the observable frame highlights the difference between the present approach of DE and others based on
usual FLRW with DE violating the strong energy condition. Using all these elements,
we have build a plausible and original DE mechanism to explain the puzzling Hubble diagram of type Ia supernovae.\\
\\
This DE mechanism presents several key features that appear to us very seducing. First, the issue of coincidence
is considerably improved. Indeed, the competition between AWE and ordinary matter cannot have occured during the radiation-dominated
era because radiation is ruled by WEP (a remark on early radiation-dominated era, where the AWE particles were relativistic, will be made
further). During the matter-dominated era, the evolution of the scalar field brings it naturally to coincidence at low redshifts ($\tilde{z}\approx 0$).
This depends crucially on the amplitude
of the non-minimal couplings to the scalar field and the cosmic coincidence can be explained when these couplings are of order unity. This is precisely
a natural consequence of the non-minimally coupled assumption, namely that the couplings to the scalar gravitational degrees-of-freedom are
of the same order as those of the tensorial ones. \\
\\
In second, we have shown that there is no need to invoke well-shaped sophisticated 
functions to justify the late cosmic acceleration. Through the calibration of three simple models, we have therefore illustrated the
robustness of the AWE DM assumption with respect to a change of parametrization.
Future works should focus on performing a compared statistical analysis of different data sets to provide
more constraints on the model parameters and more insight on the nature of AWE. 
\\
\\
Third, it should be reminded that the AWE hypothesis does not require to invoke very negative pressures
and a violation of the SEC to build a plausible DE mechanism. Cosmic acceleration is provided by the non-standard (not FLRW) terms in
the observable frame (see also \cite{fuzfa} for another models implementing this feature). Another interesting issue
that is addressed by this model is the fate of the universe. As the final state of cosmology is described by GR with
a different gravitational coupling, the final expansion regime is the one of Einstein-de Sitter (in flat cosmologies).
There is no danger of ultimate Big Rip (although the effective equation of state might be phantom) nor the cold eternity of
non-trivial vacuum-dominated de Sitter cosmology. At the opposite, the present DE mechanism is transient
and is composed of a succession of decelerating and accelerating phases while the universe relaxes to its asymptotic Einstein-de Sitter state.
\\
\\
Finally, AWE DM accounts fairly for the Hubble diagram of type Ia supernovae with
a set of cosmological parameters that is remarkably close to the predictions of the concordance model: $\tilde{\Omega}_{m,0}=0.04$,
$\tilde{\Omega}_{awe,0}=0.26$ and $\tilde{\Omega}_{\varphi,0}=0.7$. This is therefore very tempting to identify
the AWE sector to cold dark matter and the usual matter sector to baryons...
This conclusion, which should be supported by a more complete statistical analysis, has strong impacts on cosmology.
First, this gives a natural explanation of the adequacy of the concordance model.
Then, the AWE DM assumption would allow to measure the distribution of dark matter and baryons directly from supernovae data alone.
As well, the physics leading to the angular fluctuations measured in the CMB and
the gravitational (weak-)lensing of background objects by large-scale dark matter structures will be
different due to the violation of the WEP by dark matter. In the same time, the large-scale structure formation and
its impact on the local validity of GR as well as on the universality of free fall at large scales are crucial issues to be examined
for this assumption (see \cite{bertolami} for a recent interesting work
on this issue). As well, impact on fifth force constraints or local deviations from general relativity should be examined in the AWE framework,
while keeping in mind the interesting results of the recent work \cite{guendelman} on such tests with non-minimally coupled scalar fields.
Another interesting perspective is the study of the cosmic expansion in the very early universe, when the AWE DM particles
were relativistic, as an analogous mechanism of acceleration could generate inflation. This inflation would only be due
to a violation of the WEP and not of the SEC as usually assumed. \\
\\
The AWE hypothesis applied to dark matter therefore opens new interesting possibilities not only for the problem
of DE but also for its links with dark matter and inflation, i.e. for the major issues of modern cosmology.
A careful study of inflation, CMB physics and structure formation in the
framework presented here
could either constrain the coupling functions that rule the equivalence principle or rule out the AWE DM hypothesis. In the last case,
this would constitute an interesting argument in favour of a violation of SEC by DE. But, if the AWE DM appears in agreement with all other
cosmological tests, or allows to unify different aspects of cosmology, this could constitute a true step forward for this science.
This explanation of DE offers new perspectives for fundamental physics as it glimpses on the links between microphysics and gravitation as well as 
some improvements of the coincidence problem and a possible explanation of the concordance model adequacy. 
But most of all, the ideas presented in this paper allow to reduce DE to a new property 
of dark matter, a property that could potentially change our present understanding of gravitation, cosmic acceleration, 
CMB, structure formation and possibly inflation: the \textit{anomalous gravity} of dark matter.

\begin{acknowledgments}
The authors are very grateful to V. Boucher, P.-S. Corasaniti, J.-M. G\'erard, J. Larena and C. Ringeval for deep and fruitful discussions on
dark matter, dark energy and the equivalence principle. A.F. is supported by a post-doctoral fellowship from the Belgian "Fonds National de la Recherche Scientifique" (F.N.R.S.). 
A.F. also dedicates to his newly born son Hugo
these results that were partly obtained in the moments preceeding his arrival. 
\end{acknowledgments}

\appendix*
\section{Convergence Toward General Relativity Revisited}
\begin{figure}
\includegraphics[scale=0.5]{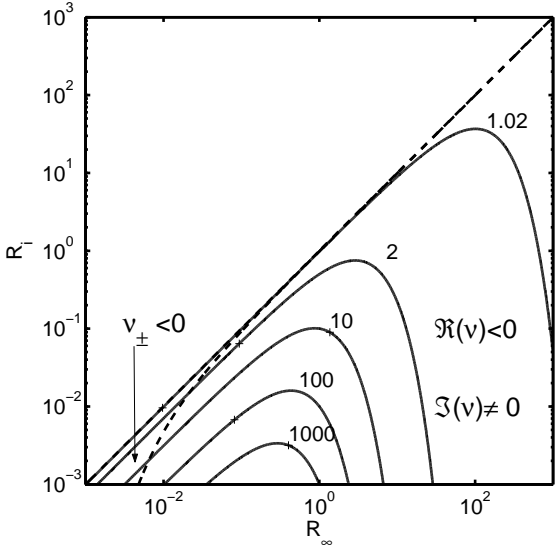}%
\caption{Asymptotic value of the gravitational coupling constant $G_c(\varphi_\infty)$ (in units of $G_*$) in the 
$(R_\infty,\;R_i)$ plane for the model with two parabolic coupling functions. The two regimes for the fixed point are separated by a dashed line. \label{fig10}}
\end{figure}

\begin{figure}
\includegraphics[scale=0.5]{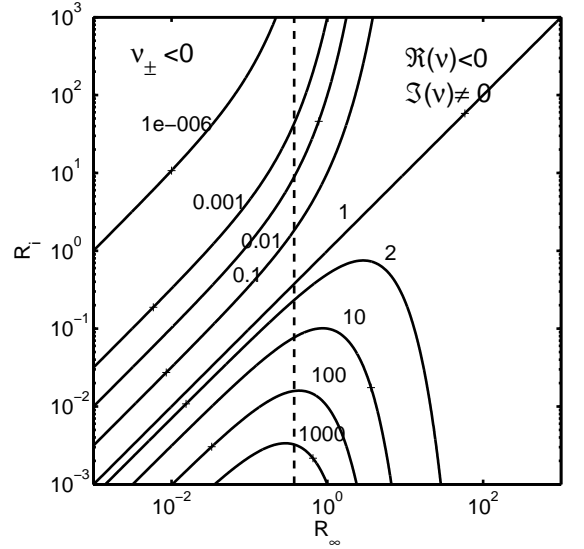}%
\caption{Asymptotic value of the gravitational coupling constant $G_c(\varphi_\infty)$ (in units of $G_*$) in the 
$(R_\infty,\;R_i)$ plane for model (\ref{model1}) ($k_m=1$). The two regimes for the fixed point, the 
in-spiralling regime ($\Re(\nu)<0$, $\Im(\nu)\neq 0$)
and the focus one ($\nu_\pm<0$) are separated by a dashed line. \label{fig11}}
\end{figure}

\begin{figure}
\includegraphics[scale=0.5]{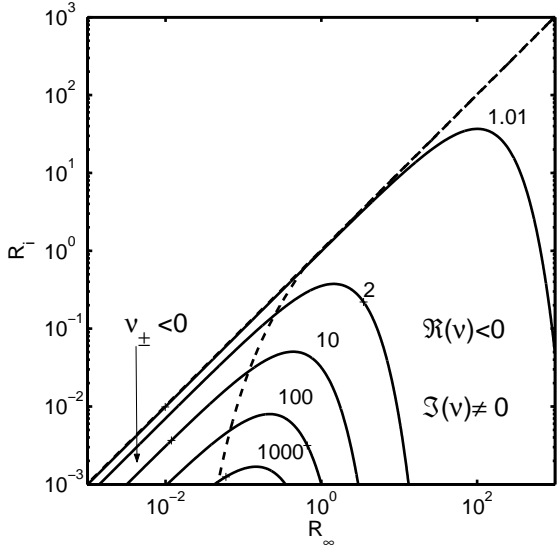}%
\caption{Asymptotic value of the gravitational coupling constant $G_c(\varphi_\infty)$ (in units of $G_*$) in the 
$(R_\infty,\;R_i)$ plane for model (\ref{model2}). The two regimes for the fixed point are separated by a dashed line. \label{fig12}}
\end{figure}

In order to characterize the fixed points (\ref{phi_inf}) of (\ref{kg2}), we can perform a linear stability analysis
for a perturbation $(\delta\varphi,\delta\varphi')$ around $(\varphi_\infty,\varphi'_\infty)$. Performing a first order 
Taylor expansion around the fixed point $(\varphi_\infty,\varphi'_\infty)$ gives
\begin{equation}
\left(
\begin{tabular}{c}
$\delta\varphi'$\\
$\delta\varphi''$
\end{tabular}
\right)=J|_{(\varphi_\infty,\varphi'_\infty)}
\left(
\begin{tabular}{c}
$\delta\varphi$\\
$\delta\varphi'$
\end{tabular}
\right)\nonumber
\end{equation}
where $J$ is the Jacobian matrix whose eigenvalues at the fixed points allow to characterize them. 
These eigenvalues are given by
\begin{equation}
\nu_\pm=\frac{3}{4}\left(-1\pm\sqrt{1-\frac{8}{3}K_\infty}\right)
\label{lambda}
\end{equation}
with $K_\infty=d\aleph(\varphi)/d\varphi|_{\varphi=\varphi_\infty}$, or in terms of the coupling functions $\alpha_m(\varphi)$ and $\alpha_{awe}(\varphi)$:
\begin{eqnarray}
K_\infty&=&\left[\frac{d\alpha_m}{d\varphi}-\alpha_{awe}\alpha_m
+\frac{\alpha_m}{\alpha_m-\alpha_{awe}}\times\right.\nonumber\\
&&\left.
\left(\frac{d\alpha_{awe}}{d\varphi}-\frac{d\alpha_{m}}{d\varphi}\right)\right]|_{\varphi=\varphi_\infty}\cdot
\label{kinf}
\end{eqnarray}
The usual case with WEP can be retrieved when $\alpha_m=\alpha_{awe}$, which gives for the position of the attractor (\ref{phi_inf})
$$
\alpha_m(\varphi_\infty)=0
$$
and for the eigenvalues parameter (\ref{kinf})
$$
K_\infty=\frac{d\alpha_m}{d\varphi}|_{\varphi=\varphi_\infty}\cdot
$$
For $1-8K_\infty/3<0$, the eigenvalues have a non-vanishing imaginary part together with a negative real part and this case corresponds to a damped oscillatory
regime of an in-spiralling stable point. For $0<1-8K_\infty/3<1$, we face a stable proper node with $\nu_\pm<0$ while the case of the hyperbolic fixed point 
$1-8K_\infty/3>1$ ($\nu_+>0$; $\nu_-<0$)
would require a negative $d\alpha_m/d\varphi$ for which the convergence toward GR at small scales would not be possible. The critically damped solution corresponds to $1-8K_\infty/3=0$.
The linear stability analysis around the fixed point $(\varphi_\infty,0)$ therefore leads to the same behavior than the non-relativistic solution of the well-known case 
of the parabolic coupling function
$\alpha_m=k_m\varphi$ (see \cite{convts}). \\
\\
Consequently, as $K_\infty >0$ ($1-8K_\infty/3<1$), the fixed point is an attractor
and the final state of AWE DM cosmology is a gravitational theory similar
to GR but defined for the value $\tilde{G}_c(\varphi_\infty)$ of the gravitational coupling constant. Quite surprisingly, the convergence mechanism toward GR 
survives despite the presence of AWE but it is now shifted to different values of the gravitational coupling constant $\tilde{G}_c$.
\\
\\
We can now go on with the non-relativistic limit $\varphi^{'2}\ll 3$ around the attractor $\varphi_\infty$ where we can approximate
the bottom of the potential well by $\aleph(\varphi)|_{\varphi\approx\varphi_\infty}\approx K_\infty \left(\varphi-\varphi_\infty\right)$ 
(see also \cite{convts}). 
Starting at some value of $\lambda=\lambda_0$ with $\varphi_0$ and $\varphi'_0$ as
initial conditions for the scalar field and its velocity, we can write down for the sub-critical motion ($0<K_\infty<3/8$):
\begin{equation}
\varphi(\lambda)=A^+e^{\nu_+ (\lambda-\lambda_0)}+A^- e^{\nu_- (\lambda-\lambda_0)}+\varphi_\infty,
\end{equation}
with 
$$
A^{\pm}=\pm\frac{\varphi'_0+\nu_{\mp}(\varphi_\infty-\varphi_0)}{\nu_+-\nu_-}
$$
where $\nu_\pm$ is given by (\ref{lambda}).
For $K_\infty>3/8$, we face a damped oscillatory regime for which the solution is
\begin{equation}
\label{phi_osc}
\varphi(\lambda)=A\exp\left(-\frac{3}{4}\lambda\right)\sin\left(\nu \lambda+\theta \right) +\varphi_\infty,
\end{equation}
with 
\begin{equation}
\nu=\frac{3}{4}\left(\frac{8}{3}K_\infty-1\right)^{1/2},
\label{freq}
\end{equation}
$$
A=\frac{\left(\varphi_0-\varphi_\infty\right)\exp\left(\frac{3}{4}p_0\right)}{\sin(\nu p_0+\theta)},
$$
and
$$
\theta=\arctan\left(\frac{\sqrt{\frac{8}{3}K_\infty-1}}{1+\frac{4\varphi'_0}{3(\varphi_0-\varphi_\infty)}}\right)-\nu p_0\cdot
$$
The critically damped solution is obtained for $K_\infty=3/8$:
\begin{equation}
\varphi(\lambda)=\frac{\varphi_0-\varphi_\infty}{1+\frac{3}{4}\lambda_0}\left(1+\frac{3}{4}\lambda\right)e^{-\frac{3}{4}(\lambda-\lambda_0)}+\varphi_\infty
\end{equation}
and 
$$
\varphi'_0=-\frac{9}{16}\lambda_0\frac{\varphi_0-\varphi_\infty}{1+\frac{3}{4}\lambda_0}
$$
\\
\\
Let us now particularize this study to different sets of the coupling functions $\alpha_m$ and $\alpha_{awe}$.
Instead of expressing these models in terms of the free parameters $(R_i,k_{awe})$ (where,
in this case, $R_i$ is the ratio between the energy densities
of usual
matter and AWE both at $\varphi=0$ at which we will start),
we prefer express the dynamics in terms of $R_\infty$ given by (\ref{rinf}).
The first model we propose is the one presented in section IV (\ref{model3}).
With this parametrization, the position of the attractor (\ref{phi_inf}) is given by:
$$
\varphi_\infty=\left(\frac{2}{k_m\left(1+R_\infty\right)}\ln\left(\frac{R_\infty}{R_i}\right)\right)^{1/2},
$$
and the eigen-frequencies parameter (\ref{kinf}) by:
$$
K_\infty=k_m^2\varphi_\infty^2R_\infty
$$
and $k_{awe}=-R_\infty k_m$ is negative if we want an attraction mechanism toward GR at small scales ($k_m>0$).
Figure \ref{fig10} summarizes the asymptotic behavior in this model. The GR attractor ($\tilde{G}_c(\varphi_\infty)=G_*$)
is retrieved for AWE sub-dominance $R_{i,\infty}\rightarrow\infty$ ($\rho^*_m\gg\rho^*_{awe}$) and for the WEP $R_i=R_\infty$.
In this parametrization, the center of the phase space $(\varphi,\varphi')=(0,0)$ is also a fixed-point. We can therefore compute the eigenvalue
parameter $K_0$ to be used for the eigen-frequencies (\ref{lambda}) instead of $K_\infty$:
\begin{equation}
K_0= k_m\frac{R_i-R_\infty}{1+R_i}.
\end{equation}
When $R_i<R_\infty$, $K_0$ is negative and we face an unstable fixed point at $\varphi=0$. The non-relativistic regime
corresponding to this case is given by (see also \cite{convts})
\begin{equation}
\label{phi_mon}
\varphi(\lambda)=A^+e^{\nu_+ \lambda}+A^- e^{\nu_- \lambda},
\end{equation}
where $\nu_\pm$ is given by (\ref{lambda}) with $K_0$ instead of $K_\infty$ and
with 
\begin{equation}
A_{\pm}=\frac{1}{2}\left(1\pm\left(1-\frac{8}{3}K_0\right)^{-1/2}\right)\varphi_i
\end{equation}
if we start at rest with $\varphi(a_*^i)=\varphi_i$.
\\
\\
Then, the simplest model one might propose assumes two Brans-Dicke coupling functions:
\begin{eqnarray}
\left\{
\begin{tabular}{l}
$A_m(\varphi)=\exp\left(k_m\varphi\right)$\\
\\
$A_{awe}(\varphi)=\exp\left(k_{awe}\varphi\right)$\\
\end{tabular}
\right.\cdot
\label{model1}
\end{eqnarray}
This model is purely illustrative as it intends to demonstrate the existence of the attracting value $\varphi_\infty$
even in this simplest case. 
This is also the parametrization of coupling functions used in chameleon cosmology \cite{chameleon}, and in fact model (\ref{model1}) itself has been first 
studied in a different context already in
\cite{damour3}.
The position of the attractor is given by, according to (\ref{phi_inf}),
\begin{equation}
\varphi_\infty=\frac{1}{k_m\left(1+R_\infty\right)}\ln\left(\frac{R_\infty}{R_i}\right)
\end{equation}
while the value of $K_\infty$ parametrizing the eigen-frequencies near the attractor is 
\begin{equation}
K_\infty=k_m^2R_\infty,
\end{equation}
and $k_{awe}=-R_\infty k_m$ is therefore of opposite sign of $k_m$.
Figure \ref{fig11} gives the relative deviation from GR at $\varphi=\varphi_\infty$ for this model. The Brans-Dicke 
theory is retrieved for $R_i=R_\infty$, meaning that both AWE DM and usual matter have the same scaling law, which corresponds well to no violation of WEP. The nature
of the fixed point is also indicated.
\\
\\
Finally, another very simple model can be formulated using a parabolic coupling function on one side and a Brans-Dicke on the other:
\begin{eqnarray}
\left\{
\begin{tabular}{l}
$A_m(\varphi)=\exp\left(k_m\frac{\varphi^2}{2}\right)$\\
\\
$A_{awe}(\varphi)=\exp\left(k_{awe}\varphi\right)$\\
\end{tabular}
\right.\cdot
\label{model2}
\end{eqnarray}
Once again, let us express the asymptotic quantities in terms of the initial and final ratio $R_i$ and $R_\infty$ 
to get for the position of the attractor:
\begin{equation}
\varphi_\infty=\left(\frac{2}{k_m\left(1+2R_\infty\right)}\ln\left(\frac{R_\infty}{R_i}\right)\right)^{1/2}
\end{equation}
and for $K_\infty$
\begin{equation}
K_\infty=k_m+k_m^2R_\infty\varphi_\infty^2-\frac{k_m}{1+R_\infty}
\end{equation}
and $k_{awe}=-k_m R_\infty \varphi_\infty$.
Figure \ref{fig12} gives the absolute deviation from GR at $\varphi=\varphi_\infty$ for this model. The GR attractor
is retrieved for AWE sub-dominance $R_{i,\infty}\rightarrow\infty$ ($\rho^*_m\gg\rho^*_{awe}$) and for the weak equivalence
principle $R_i=R_\infty$. Models (\ref{model3}), (\ref{model1}) and (\ref{model2}) illustrate that the attraction mechanism toward $\varphi_\infty$ exists for various choices of 
the couplings to the scalar field, even in the most simple cases.
\begin{figure}
\includegraphics[scale=0.4]{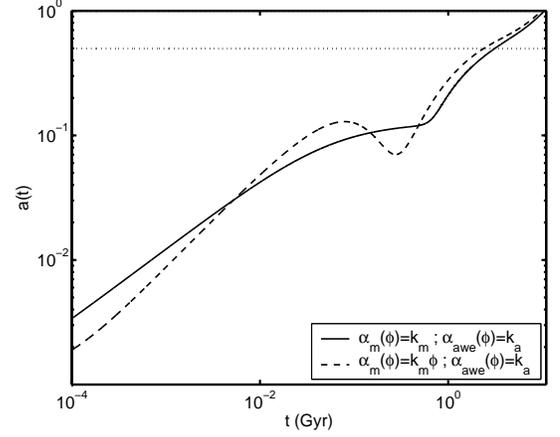}
\caption{Cosmic expansion $\tilde{a}(\tilde{t})$ for the models (\ref{model1}) and (\ref{model2}) (see also Table \ref{tab1}).
The dotted horizontal line represents the redshift range of the data used for calibration \label{fig13}}
\end{figure}\\
\\
Let us now illustrate how the parametrizations (\ref{model1}) and (\ref{model2}) can be used for building a DE mechanism.
Figure \ref{fig13} reproduces the evolution of the scale factor for the calibrated models (\ref{model1}) and (\ref{model2}) (see Table I for the parameters).
In models (\ref{model1}) and (\ref{model2}), the Hubble diagram is reproduced by one oscillation of $\tilde{G}_c$
around the position of the attractor $\tilde{G}_c(\varphi_\infty)$.  
In model (\ref{model2}), the previous oscillations are so strong that they even yield an ambiguity in the definition of the cosmological redshift $\tilde{z}=\tilde{a}_0/\tilde{a}-1$
with respect to time $\tilde{t}$
around $\tilde{z}\approx 10$. However, these simple parametrizations allow faithfull reproduction of Hubble diagram data and show how the AWE DM mechanism
for dark energy is robust against a change of parametrization.


\begin{thebibliography}{99}
\bibitem{sdss} M. Tegmark et al., Phys.Rev. D74 (2006) 123507 ; \\
U. Seljak, Phys.Rev.D71 103515 (2005)\\
M. Tegmark et al., Phys.Rev.D 69 103501 (2004).
\bibitem{2DF} S. Cole et al., Mon.Not.Roy.Astron.Soc. 362 (2005) 505-534\\
G. Efstathiou et al., Mon.Not.Roy.Astron.Soc. 330 (2002) L29.
\bibitem{cobe} J.C. Mather et al., ApJ 354, L37 (1990)\\
G.F. Smoot et al., ApJ 371, L1 (1991)\\
G.F. Smoot et al., ApJ 396, L1 (1992)\\
E.L. Wright, ApJ 396, L13 (1992)\\
K.M. Gorski et al., ApJ 464, L11 (1996).
\bibitem{boomerang} P. de Bernardis et al., Nature 404, 955 (2000).
\bibitem{wmap} D.N. Spergel, et al., 2003, ApJS, 148, 175\\
D.N. Spergel et al., ApJ, in press, astro-ph/0603451
\bibitem{snls} P. Astier et al.,  Astron. Astrophys. 447 31 (2006)
\bibitem{scp} S. Perlmutter et al., ApJ 483, 565 (1997)\\
S. Perlmutter et al., Nature 391, 51 (1998)\\
S. Perlmutter et al., ApJ 517, 565 (1999)\\
R. A. Knop et al, ApJ 598:102-137, 2003
\bibitem{highz} P. Garnavich  et al. , 1998, ApJ, 509, 74;\\
A. Riess et al., Astron. J. 116, 1009 (1998)
A. Riess et al., ApJ 560, 49 (2001).
\bibitem{fuzfa} A. F\"uzfa, J.-M. Alimi, Phys. Rev. Lett. 97, 061301 (2006);\\
A. F\"uzfa, J.-M. Alimi, Phys.Rev. D73 (2006) 023520;\\
J.-M. Alimi, A. F\"uzfa, Proceedings of the International Workshop "From Quantum to Cosmos: Fundamental Physics Research in Space",
held in Washington, DC, USA, May 2006;\\
A. F\"uzfa, J.-M. Alimi, Proceedings of the SF2A 2006 Conference, held in Paris, France, June 2006, astro-ph/0609099;\\
A. F\"uzfa, J.-M. Alimi, Proceedings of the 11th Marcel Grossmann Conference, held in Berlin, Germany, July 2006, gr-qc/0702085
\bibitem{pscora} L. Amendola, Phys.Rev. D62 (2000) 043511, astro-ph/9908023; 
S. Das, P.-S. Corasaniti, J. Khoury, Phys.Rev. D73 (2006) 083509
\bibitem{caroll} Carroll, Sean M. (2004). Spacetime and Geometry: An Introduction
to General Relativity, San Francisco, Addison-Wesley
\bibitem{chameleon} J. Khoury, A. Weltman, Phys. Rev. Lett. 93, 171104 (2004);\\
J. Khoury, A. Weltman, Phys. Rev. D69, 044026 (2004);\\ P. Brax, C. van de Bruck, A.C. Davis, J. Khoury, A. Weltman, Phys. Rev. D 70, 123518
(2004)\\
D. F. Mota \& D. J. Shaw, Phys. Rev. Lett. 97 151102 (2006);\\
D. F. Mota \& D. J. Shaw, Phys. Rev. D 75, 063501 (2007), hep-ph/0608078
\bibitem{einstein} A. Einstein, "\textit{Cosmological Considerations on the General Theory of Relativity}", in "\textit{The Principle of Relativity}", Dover, 1952.
\bibitem{buchert} T. Buchert, J. Larena \& J.-M. Alimi, Class.Quant.Grav. 23 (2006) 6379-6408.
\bibitem{weinberg} S. Weinberg, Rev. Mod. Phys. 61, 1 (1989)
\bibitem{peebles} B. Ratra  \& P.J.E. Peebles, Phys. Rev. D37 (1988) 3406\\
C. Wetterich, Nucl. Phys. B 302, 668 (1988)
\bibitem{steinhardt}   Steinhardt P.J., Wang L. \& Zlatev I., Phys.Rev. D59 (1999) 123504,
astro-ph/9812313. \\
Zlatev I., Wang L. \& Steinhardt P.J., Phys.Rev.Lett. 82 (1999)
896-899, astro-ph/9807002.
\bibitem{brax} P. Brax \& J. Martin, JCAP 11 (2006), 008.
\bibitem{ts} P. Jordan, Nature 164, 637 (1949)\\
M. Fierz, Helv. Phys. Acta 29, 128 (1956)
C. Brans and R. H. Dicke, Phys. Rev. 124, 925 (1961)
\bibitem{convts} A. Serna, J.-M. Alimi and A. Navarro, Class.Quant.Grav. 19 (2002) 857-874;\\
A. Serna and J.-M. Alimi, Phys.Rev. D53, 3074 (1996);\\
T. Damour and K. Nordtvedt, Phys. Rev. D48 (8),
  3436-3450 (1993);\\
T. Damour and K. Nordtvedt, Phys. Rev. Lett. 70 (15),
  2217-2219 (1993)\\
J.D. Barrow, Phys. Rev. D 47, 5329 - 5335 (1993)
\bibitem{extq}  J.-P. Uzan, Phys. Rev. D 59, 123510 (1999)\\
L. Amendola, Phys. Rev. D 60, 043501 (1999) \\
F. Perrotta, C. Baccigalupi \& S. Matarrese, Phys. Rev. D 61, 023507 (1999)
\bibitem{massd} C.T. Will \& G.C. Ross, Nucl. Phys. B 311, 253 (1988);\\
J. Ellis, S. Kalara, K.A. Olive \& C. Wetterich, Phys. Lett. B 228, 264 (1989);\\
C. Wetterich, Astron. Astrophys. 301, 321 (1995);\\
G.W. Anderson \& S.M. Caroll, astro-ph/9711288;\\
G. Huey, P.J. Steinhardt, B.A. Ovrut \& D. Waldram, Phys. Lett. B 476, 379 (2000);\\
D. F. Mota, J. D. Barrow, Mon. Not. Roy. Astron. Soc. 349, 291 (2004), astro-ph/0309273\\
D. F. Mota, J. D. Barrow, Phys.Lett.B 581 141-146 (2004), astro-ph/0306047
\bibitem{damour2} T. Damour, D. Polyakov, Nucl.Phys. B423 (1994) 532-558.\\
T. Damour, D. Polyakov, Gen. Rel. Grav. 26, 1171 (1994)
\bibitem{bertolami} O. Bertolami, F. Gil Pedro \& M. Le Delliou, astro-ph/0703462.
\bibitem{guendelman} E.I. Guendelman \& A.B. Kaganovich, arXiv:0704.1998.
\bibitem{damour3} T. Damour, G.W. Gibbons \& C. Gundlach, Phys. Rev. Lett. 64, 123 (1990)

\end{thebibliography}

\end{document}